\documentclass[runningheads]{llncs}
\usepackage[T1]{fontenc}
\usepackage{graphicx}
\usepackage{amsmath,amssymb,amsfonts}
\usepackage{xcolor}
\usepackage{verbatim}
\usepackage{bbding}
\usepackage{multirow}
\usepackage{booktabs}
\usepackage{makeidx}
\makeindex
\usepackage{enumitem}
\setlist[itemize]{noitemsep, topsep=2pt}

\newcommand{\topup}{$\mathrm{topup}$}

\newcommand{\mypar}[1]{\noindent\textbf{#1}\,}

\begin{document}
\title{Susceptibility Distortion\index{susceptibility distortion} Correction of Diffusion MRI\index{Diffusion MRI} with a single Phase-Encoding\index{Phase Encoding} Direction}
%
%
\titlerunning{Susceptibility Distortion\index{susceptibility distortion} Correction of dMRI\index{dMRI} with a Single PE Direction}

\author{
Sedigheh Dargahi\inst{1} \and
Sylvain Bouix\inst{1}\Envelope \and
Christian Desrosiers\inst{1}
}

\authorrunning{S. Dargahi et al.}

\institute{
École de technologie supérieure, Montreal, QC, Canada \\
\email{sedigheh.dargahi.1@ens.etsmtl.ca} \\
\email{sylvain.bouix@etsmtl.ca} \\
\email{christian.desrosiers@etsmtl.ca}
}

\maketitle              
\begin{abstract}
Diffusion MRI (dMRI)\index{dMRI}\index{Diffusion MRI} is a valuable tool to map brain microstructure and connectivity by analyzing water molecule diffusion in tissue. However, acquiring dMRI\index{dMRI} data requires to capture multiple 3D brain volumes in a short time, often leading to trade-offs in image quality. One challenging artifact is susceptibility-induced distortion\index{susceptibility distortion}, which introduces significant geometric and intensity deformations. Traditional correction methods, such as \topup{}\index{topup}, rely on having access to blip-up\index{blip-up} and blip-down\index{blip-down} image pairs, limiting their applicability to retrospective data acquired with a single phase encoding\index{Phase Encoding} direction. In this work, we propose a deep learning-based\index{deep learning} approach to correct susceptibility distortions\index{susceptibility distortion} using only a single acquisition (either blip-up\index{blip-up} or blip-down\index{blip-down}), eliminating the need for paired acquisitions. Experimental results show that our method achieves performance comparable to \topup{}\index{topup}, demonstrating its potential as an efficient and practical alternative for susceptibility distortion\index{susceptibility distortion} correction in dMRI\index{dMRI}.

\end{abstract}

\section{Introduction}
Among neuroimaging techniques, diffusion MRI (dMRI)\index{dMRI}\index{Diffusion MRI} plays a crucial role in understanding the complex connectivity and microstructural features of the brain \cite{book1}. The acquisition of dMRI\index{dMRI} data involves two key considerations. First, special gradient fields are applied during data collection to make the image sensitive to diffusion in a specific direction. Second, to acquire the needed information, many separate 3D images of the brain must be obtained, as each 3D image provides information about diffusion in just one direction. Consequently, to acquire dMRI\index{dMRI} data in a reasonable amount of time--typically a few minutes--fast imaging sequences such as Echo Planar Imaging (EPI)\index{echo planar imaging} are used. Unfortunately, these fast acquisitions often lead to image artifacts, which need to be addressed in post-processing. One of the main artifacts, known as \emph{susceptibility distortion}\index{susceptibility distortion}, changes the geometry of the brain along the phase encoding (PE)\index{phase encoding} direction. 

When dealing with susceptibility-induced distortions\index{susceptibility distortion}, both traditional and deep learning (DL)\index{deep learning} techniques have been developed. One of the popular strategies is to acquire reversed-phase encoding\index{Phase Encoding} directions from which a field map can be estimated (the \topup{}\index{topup} tool in FSL)\index{topup} \cite{ref_article2}. In recent years, several deep learning-based\index{deep learning} susceptibility distortion\index{susceptibility distortion} correction techniques \cite{ref_article1,ref_article8} have emerged to correct this kind of distortion faster and more accurately than \topup{}\index{topup} \cite{survey1}.  In \cite{ref_article3}, an unsupervised U-Net\index{U-Net} minimizes the difference between unwarped images at multiple resolutions to accelerate processing.  The approach in \cite{ref_article4} uses fiber orientation distributions (FODs)\index{fibre orientation distribution (FOD)} derived from dual-phase dMRI\index{dMRI} and applies a U-Net (DrC-Net) for correction. FOD estimation is computationally intensive, as it requires fitting high-order spherical harmonic models at each voxel to resolve crossing fibers. When applied to millions of voxels in multi-shell diffusion data, this results in substantial computational demands. \cite{ref_article5} employs PSF-EPI images as ground truth for training a correction network, yet this data is rarely acquired in clinical practice.

A common challenge of these methods is the dependence on blip-up\index{blip-up} and blip-down\index{blip-down} acquisitions, which are not always available. This reliance limits the broader use of advanced distortion correction techniques. Developing a method that works across different scenarios and data types would thus make distortion correction more accessible and practical for both clinical use and research settings. One of the promising DL method \cite{ref_article1} addresses this problem by synthesizing an undistorted EPI image from a structural T1-Weighted (T1w) scan and a single-blip diffusion image. The single-blip image and synthetized image are then used as input to \topup{}\index{topup} to estimate a field map. However, training this method relies on having access to a dataset containing undistorted multi-shot diffusion b0 images paired with single-blip data, which are not commonly acquired. Additionally, \topup{}\index{topup} is still needed to estimate the field map.

In this work, we introduce a slice-wise deep-learning\index{deep learning} method designed to correct susceptibility distortions\index{susceptibility distortion} in dMRI\index{dMRI} using only \emph{a single phase-encoding direction}\index{Phase Encoding}, addressing a limitation in current distortion correction techniques. Our contributions can be summarized as follows. First, unlike traditional approaches such as \cite{ref_article2}, which require paired acquisitions, our method needs only one phase-encoded\index{Phase Encoding} distorted dMRI\index{dMRI} image alongside a corresponding structural T1w image. This broadens the practical applicability, particularly for retrospective datasets where dual-phase acquisitions are not available. Second, our model simultaneously predicts both the Voxel Displacement Map (VDM)\index{voxel displacement map (VDM)} and the intensity-corrected b0 image in a single forward pass, simplifying the correction workflow. Finally, by integrating these ideas, we reduce processing times from several minutes typically needed by well-known methods such as Synb0 \cite{ref_article1} to mere seconds, facilitating large-scale studies and time-sensitive applications. Experimental results demonstrate that our method not only provides correction quality close to that of dual-phase methods like topup\index{topup}, but also surpasses existing single-phase technique in both accuracy and speed.

\section{Method}
\begin{figure}[!t]
\includegraphics[width=\textwidth]{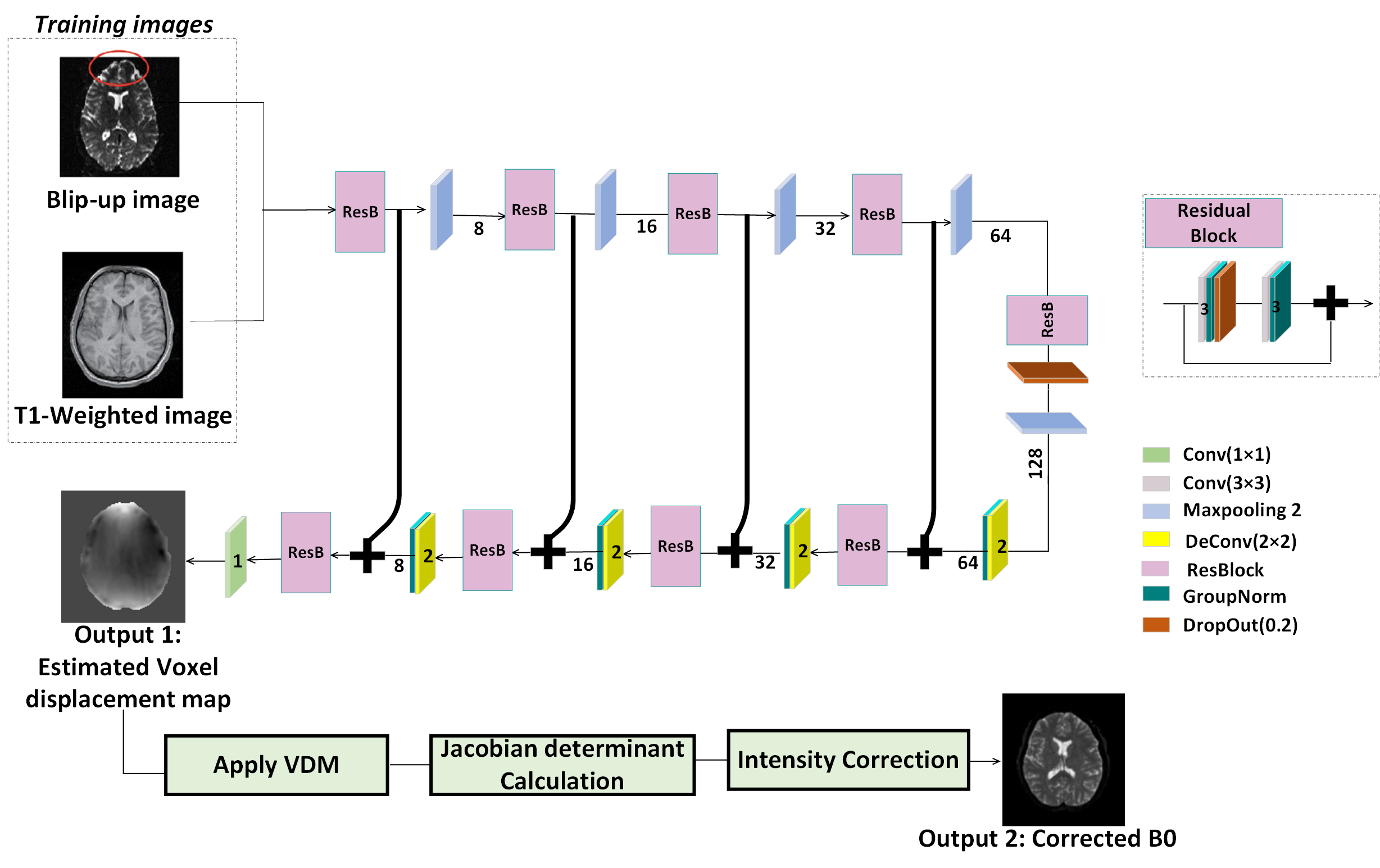}
        \caption{Architecture of the proposed method.}  \label{fig1}
\end{figure}

As illustrated in Figure \ref{fig1}, the proposed model takes as input distorted blip-up\index{blip-up} or blip-down\index{blip-down} b0 images\footnotemark{} as well as T1w images, and outputs the voxel displacement map for correcting the b0 image. The following sections detail the network architecture, loss function and preprocessing pipeline of our proposed method.
\footnotetext{Without loss of generality, we assume in the rest of the paper that blip-up images are given.}

\subsection{Model Architecture}

Our proposed model builds on a U-Net architecture, as illustrated in Figure~\ref{fig1}. The encoder consists of sequential residual blocks, each using \(3 \times 3\) convolutional kernels and skip connections to maintain gradient flow and encourage stable learning. As we progress deeper into the encoder, the number of feature maps increases from 8 to 128, with downsampling performed via max-pooling. The bottleneck features a single residual block with 128 channels, followed by a dropout layer with a rate of 0.2 to help prevent overfitting. The decoder mirrors the encoder's structure, using transposed convolutions to upsample features and gradually reduce the number of channels from 128 back down to 8. Skip connections between corresponding encoder and decoder layers help retain spatial details. The output is generated by a \(1 \times 1\) convolutional layer, which produces a single-channel VDM\index{voxel displacement map (VDM)}.

To further mitigate overfitting, we made several architectural and regularization choices.  Unlike the original U-Net \cite{ref_article7}, which begins with 64 channels and expands up to 1024 over five downsampling layers, our model is shallower and lighter, starting with just 8 channels and growing to 128 over four levels. To further improve generalization, we introduced dropout (with a probability of 0.2) inside every residual block and after the bottleneck, which is not part of the original U-Net design. Lastly, we apply L1 regularization to penalize large weights, encouraging sparsity in the model parameters. The L1 penalty is added to the total loss with a scaling factor \(\lambda_{\text{reg}} = 10^{-5}\).

\mypar{2.5D Convolutional Approach.}  While susceptibility distortions\index{susceptibility distortion} only occur along the phase encoding\index{Phase Encoding} direction within individual 2D slices and are inherently one-dimensional and often localized, our model adopts a 2.5D convolutional strategy to incorporate some 3D context. Instead of using computationally expensive 3D convolutions, we process each slice together with its adjacent neighbors (one slice above and one below) as a 3-channel input. For edge slices (the first and last slices), the nearest slice is duplicated to maintain the three-slice configuration. This approach allows the model to incorporate contextual information from neighboring slices without the need for a full 3D model, striking a good balance between anatomical awareness and computational efficiency.

\mypar{Distortion correction.} The proposed method produces two outputs for each input slice. The first is the predicted voxel displacement map, which estimates spatial distortions along the phase encoding\index{Phase Encoding} direction in millimeters.  The second output is the distortion-corrected b0 image, referred to as \(b0^{DL}\). To generate this corrected image, we follow a series of steps based on the procedure described in \cite{ref_article6}:

\begin{itemize}

\item \textbf{Displacement Grid Creation}: Using the VDM\index{voxel displacement map (VDM)}, a displacement grid is created, where the VDM\index{voxel displacement map (VDM)} values represent displacements along the phase encoding\index{Phase Encoding} direction. The displacement grid is applied to the first b0 imaging in the dMRI\index{dMRI} volume.

\item \textbf{Intensity Correction with the Jacobian Determinant\index{Jacobian determinant}}: To account for intensity variations, the Jacobian determinant\index{Jacobian determinant} of the displacement field is calculated as follows:
\begin{equation}
J_{\text{Field}}(x, y) \,=\, 1 + \frac{\partial \mathrm{VDM}(x, y)}{\partial y}
\end{equation}
Here $y$ represents the phase-encoding direction\index{Phase Encoding}.  The intensity of the corrected b0 image is then adjusted by multiplying it with the corresponding Jacobian determinant\index{Jacobian determinant} values:
\begin{equation}
\text{b0}^{DL}(x, y) \, =\, J_{\text{Field}}(x, y) \cdot \text{b0}_{\text{corrected}}(x, y)
\end{equation}

\item \textbf{Final Output}: The resulting image is the corrected \( \text{b0}^{DL} \), which is generated by applying this procedure slice by slice, and then stacking the corrected slices to reconstruct the full 3D volume.

\end{itemize}

\subsection{Loss Function}

The loss function employed to train the model consists of four terms: L1 loss on the VDMs\index{voxel displacement map (VDM)}, L2 loss on the gradients of the VDMs\index{voxel displacement map (VDM)}, Structural Similarity Index Measure (SSIM) loss on the corrected b0 images, and Mutual Information (MI) between the T1w and the corrected b0 images. All four losses are computed only within the brain region, defined by a binary mask derived from the T1w image and dilated by 3 pixels to avoid edge artifacts. The choice of these losses is based on the following motivations. We use an L1 loss for the predicted VDMs\index{voxel displacement map (VDM)} to reduce the impact of large errors, which frequently occur at the boundaries between different brain tissues. The gradient term is defined as the L2 of the difference between the partial derivatives of the predicted and reference VDMs\index{voxel displacement map (VDM)} along the two in-plane axes. It serves as a regularizer that constrains local variations in the deformation field, helping to prevent abrupt changes and encouraging the model to capture structural edges of VDM\index{voxel displacement map (VDM)}. To evaluate image-to-image alignment quality, we employ SSIM, as it emphasizes the preservation of structural details, particularly at region boundaries. Finally, we use MI to compare the T1w and b0 images, as it is well-suited for assessing alignment between different imaging modalities. We compute global MI by flattening the masked intensities into a 32-bin joint histogram, then smooth it with a separable Gaussian kernel (\(\sigma = 1.0\)) to suppress noise and binning artifacts. The negative MI is used as the loss to encourage alignment. The overall loss function is given by
\begin{align}
\mathcal{L}_{\mathrm{total}}
&= \lambda_{1}\,\big\lVert \mathrm{VDM}^{\mathrm{topup}}
- \mathrm{VDM}^{\mathrm{DL}}\big\rVert_{1}
\;+\;\lambda_{2}\,\big\lVert \nabla\,\mathrm{VDM}^{\mathrm{topup}}
- \nabla\,\mathrm{VDM}^{\mathrm{DL}}\big\rVert_{2}
\nonumber \\
&\quad +\;\lambda_{3}\,\mathrm{SSIM}\bigl(\mathrm{b0}^{\mathrm{topup}},
\,\mathrm{b0}^{\mathrm{DL}}\bigr)
\;+\;\lambda_{4}\,\mathrm{MI}\bigl(\mathrm{T1w},\,\mathrm{b0}^{\mathrm{DL}}\bigr)
\end{align}
We place the greatest emphasis on the VDM\index{voxel displacement map (VDM)} prediction by setting the L1 loss weight \(\lambda_1\) to 1, and choose weights of \(\lambda_2=0.5\), \(\lambda_3=0.3\), and \(\lambda_4=0.5\) for the gradient, SSIM, and MI terms, respectively.

\subsection{Preprocessing Pipeline}

The preprocessing steps are designed to align and prepare the data for input into the model. At first, brain masks are generated for both T1w images and b0 volumes to ensure the focus is on brain tissue and to exclude irrelevant regions. We generate the b0 brain masks using FSL’s BET, and the T1w images are masked using FastSurfer’s segmentation outputs. Then, the T1w images are rigidly registered to the b0 volumes and resampled to match the dimensions of the b0 images. Finally, three consecutive slices of b0 images are concatenated with the corresponding three slices of T1w images, resulting in a six-channel input with dimensions \(6\!\times\!128\!\times\!128\!\times\!80\).

\section{Experiments}

\mypar{Dataset.} To tackle susceptibility distortions\index{susceptibility distortion} in dMRI\index{dMRI}, our approach only relies on a single phase-encoding\index{Phase Encoding} dMRI\index{dMRI}. We also require a T1w image to provide high-resolution structural information, helping the model guide what an undistorted brain should look like. The dataset used for training is the National Institute of Mental Health Intramural Healthy Volunteer (NIMH-HV) Dataset\index{datasets!NIMH-HV} \cite{ref_article9}, which is available on OpenNeuro (https://doi.org/10.18112/openneuro.ds005752.v2.1.0). This dataset includes both blip-up\index{blip-up} and blip-down\index{blip-down} acquisitions, allowing us to use dual-phase \topup{}\index{topup} corrections as our ground truth reference for training. The dMRI\index{dMRI} scans were acquired with the following parameters: echo time (TE) of 60.7 ms, repetition time (TR) of 7.8 s, and voxel dimensions of 1.8125×1.8125×2.0 mm³. Each scan includes 6 non-diffusion-weighted (b0) volumes and 48 diffusion-weighted directions. This data comprises 125 samples after preprocessing (as not all subjects have DWI). The shape of the data for each subject is \(128 \times 128 \times 80 \times 54\). 
To evaluate how well the model generalizes, we also test it on unseen data of compressed-sensing diffusion spectrum imaging (CS-DSI)\index{datasets!CS-DSI} \cite{ref_article10} which has 20 subjects available on OpenNeuro (https://doi.org/10.18112/openneuro.ds004737.v2.0.0). CS-DSI dMRI\index{dMRI} scans were acquired with the following parameters: echo time (TE) of 0.09 s, repetition time (TR) of 4.3 s, and voxel dimensions of 1.691×1.691×1.7 mm³. Each scan includes 7 b0 volumes and 1 diffusion-weighted direction. 

\vspace{0.5em}
\noindent We use the following in our approach:
\begin{itemize}
    \item \textbf{Input Data}: The b0 volumes from the DWI dataset are selected as they serve as a reference image free from the diffusion-weighted gradients applied during dMRI\index{dMRI} scans. These images have a shape of \(128 \times 128 \times 80\), corresponding to 80 slices per volume.
   \item  \textbf{Silver Standard}: We generate our reference distortion field and corrected b0 image using FSL's \topup{}\index{topup} on paired blip-up/blip-down\index{blip-up}\index{blip-down} b0 volumes.
    We convert the field map in units of Hz to a voxel displacement map in millimeters using the read-out time and the phase-encode\index{Phase Encoding} voxel size \cite{ref_article6}:
    \begin{equation}
    \mathrm{VDM}^{\mathrm{topup}}(x, y)
      \,= \mathrm{FM}(x, y)
      \;\times\;\mathrm{Read\_Out\_Time}
      \;\times\;\mathrm{VoxelSize}_y
    \end{equation}
    Finally, to match our network's outputs, we re-apply this VDM\index{voxel displacement map (VDM)} (and its Jacobian determinant\index{Jacobian determinant} for intensity modulation) to the original distorted b0, exactly as described in \cite{ref_article6}. The resulting silver‐standard $(\mathrm{VDM}^{\mathrm{topup}},\,\mathrm{b0}^{\mathrm{topup}})$ is used to supervise training of our model.
\end{itemize}

\mypar{Orientation Focus and Data Split.} Susceptibility distortions\index{susceptibility distortion} occur along the phase encoding\index{Phase Encoding} direction of the 2D acquisition plane. In most datasets, including the NIMH-HV, the acquisition plane is typically the axial plane and the phase encoding\index{Phase Encoding} is along the anterior-posterior axis, which is the scenario we assume for our current model. 

In our configuration, the NIMH-HV dataset is divided as follows: 75\%  of the data are assigned for training which would be 93 subjects, 15\%  of the data for validation which would be 18 subjects, and the remaining 10\%  for test phase which would be 14 subjects. The splitting is performed randomly to ensure a balanced representation of the dataset. We also applied {\em on-the-fly augmentations}, where each training slice undergoes, with 50\% probability, one of the following transformations: a random integer-valued translation of up to $\pm5$ pixels in both directions; a square crop of variable size (50\%--90\% of the image) followed by zero-padding back to the original dimensions; additive Gaussian noise ($\sigma=0.05$) to simulate acquisition variability. In addition, we included horizontal flips and a ``mixcut'' operation, in which the right half of one subject's slice is spliced with the corresponding left half of another subject's slice to further diversify spatial patterns. These two types of augmentation are well-suited for our application, as susceptibility distortion\index{susceptibility distortion} occurs only along the y-axis. As a result, the silver-standard VDMs\index{voxel displacement map (VDM)} of horizontally flipped or half-cut images can be used directly without modification. By combining these five augmentations, we encourage the network to learn distortion-invariant features while preserving anatomical consistency.

\mypar{Implementation Details.} The model is implemented in PyTorch and trained on a Linux system with an NVIDIA RTX A6000 GPU. We use the Adam optimizer with an initial learning rate of  \(10^{-3}\), and apply a scheduler that halves the learning rate if the validation loss does not improve for 5 epochs. To avoid overfitting, early stopping halts training after 30 stagnant epochs. The model is trained for up to 96 epochs with a batch size of 8, total of 62 minutes of training.

\section{Results}
\subsection{Comparison with state-of-the-art methods}

We evaluate the performance of our model against Synb0\index{Synb0}, using \topup{}\index{topup} as the reference on both the NIMH-HV and the external CS-DSI data; the results are reported in Table \ref{tab:tab1}. We chose Synb0 for comparison because, like our method, it works with only a single blip-up\index{blip-up} or blip-down\index{blip-down} image, making it a fair comparison. To ensure consistency, we took FM estimated by Synb0 and applied the same steps used for our method and \topup{}\index{topup}, following the procedure in \cite{ref_article6}, to generate both the VDM\index{voxel displacement map (VDM)} and the corrected b0 image. As shown in Table \ref{tab:tab1}, our method outperforms Synb0 in terms of both VDM\index{voxel displacement map (VDM)} and b0 Root Mean Squared Error (RMSE). On the NIMH-HV, the VDM\index{voxel displacement map (VDM)} RMSE is reduced by  about 53\%, while the b0 RMSE improves by about 8\%. A similar trend holds for the CS-DSI data, where our model reduces the VDM\index{voxel displacement map (VDM)} RMSE by roughly one‑third and achieves a slight reduction in b0 RMSE. We also compute the mutual information between each corrected b0 image and the T1w image. While \topup{}\index{topup} remains the silver standard with the highest MI, our method boosts MI relative to Synb0 (approximately 0.5\% on NIMH-HV, and 6.1\% on CS-DSI), indicating improved anatomical alignment, although these improvements are small. 

In terms of runtime, Synb0 requires separate steps to correct each volume: almost 15 minutes to synthesize the undistorted b0 image, followed by more than 6 minutes to run \topup{}\index{topup} using that synthesized image as input for correction. In contrast, our method performs the entire correction in a single forward pass and completes inference in just a few seconds, making it much faster for large-scale or time-sensitive studies. Overall, these results demonstrate that our deep learning–based\index{deep learning} correction reduces distortion errors compared to Synb0, even when tested on completely unseen CS-DSI volumes. Additionally, the runtimes improves from tens of minutes to just a few seconds.

\begin{table}[ht]
\caption{Comparison of VDM\index{voxel displacement map (VDM)} and b0 RMSE for Synb0 and our method with ground truth \topup{}\index{topup}; and Mutual Information (MI) comparison of \topup{}\index{topup}, Synb0 and our method with T1w image on NIMH-HV and CS-DSI datasets. All metrics are averaged across the held-out test subjects and reported as mean (standard deviation).}
\label{tab:tab1}
\centering
\setlength{\tabcolsep}{0.5em}
\resizebox{.92\linewidth}{!}{
\begin{tabular}{l|l|c|c|c}
\toprule
\textbf{Dataset} & \textbf{Method} & \textbf{VDM RMSE}\,$\downarrow$ & \textbf{b0 RMSE}\,$\downarrow$  & \textbf{MI}\,$\uparrow$ \\
\midrule
\multirow{3}{*}{NIMH-HV} 
    & \topup{}  & n/a             & n/a                 & 0.7620 (0.0785) \\
    & Synb0  &  2.34 (0.53)     &  $1.91 \times 10^2$ (0.348$\times 10^2$) & 0.7029 (0.1031) \\
    & Ours   &  1.10 (0.31)     &  $1.76 \times 10^2$ (0.339$\times 10^2$) & 0.7062 (0.0743) \\
\midrule
\multirow{3}{*}{CS-DSI} 
    & \topup{}  & n/a             & n/a                 & 0.5492 (0.0483) \\
    & Synb0  & 1.95 (0.26)     & $7.49 \times 10^2$ (1.19$\times 10^2$) & 0.4378 (0.0537) \\
    & Ours   & 1.31 (0.11)     & $7.46 \times 10^2$ (1.11$\times 10^2$) & 0.4645 (0.0384) \\
\bottomrule
\end{tabular}
}
\end{table}

\begin{figure}[!t]
  \centering
  \begin{minipage}[b]{0.48\textwidth}
    \centering
    \fbox{\includegraphics[width=\textwidth]{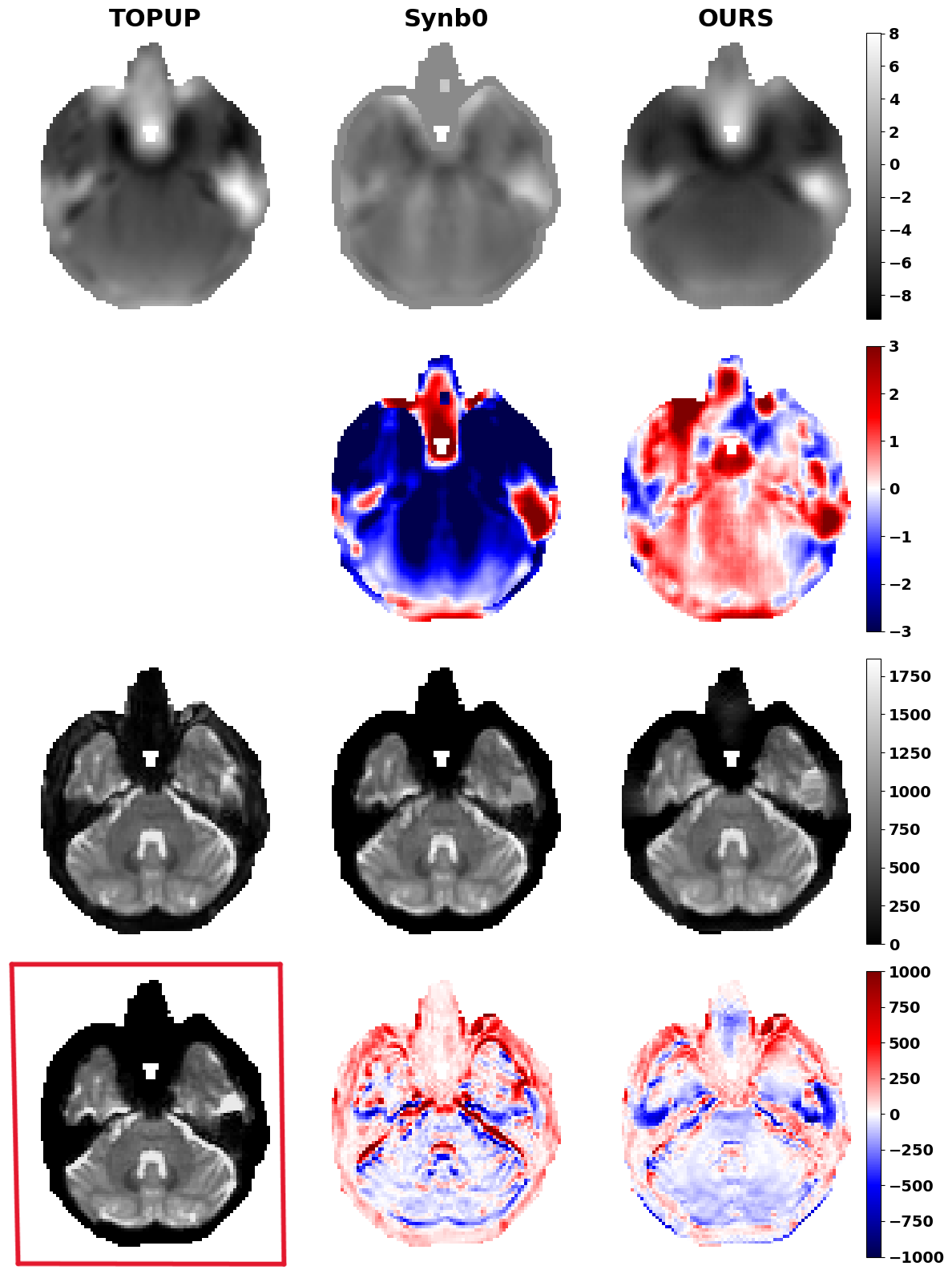}}
  \end{minipage}
  \hfill
  \begin{minipage}[b]{0.48\textwidth}
    \centering
    \fbox{\includegraphics[width=\textwidth]{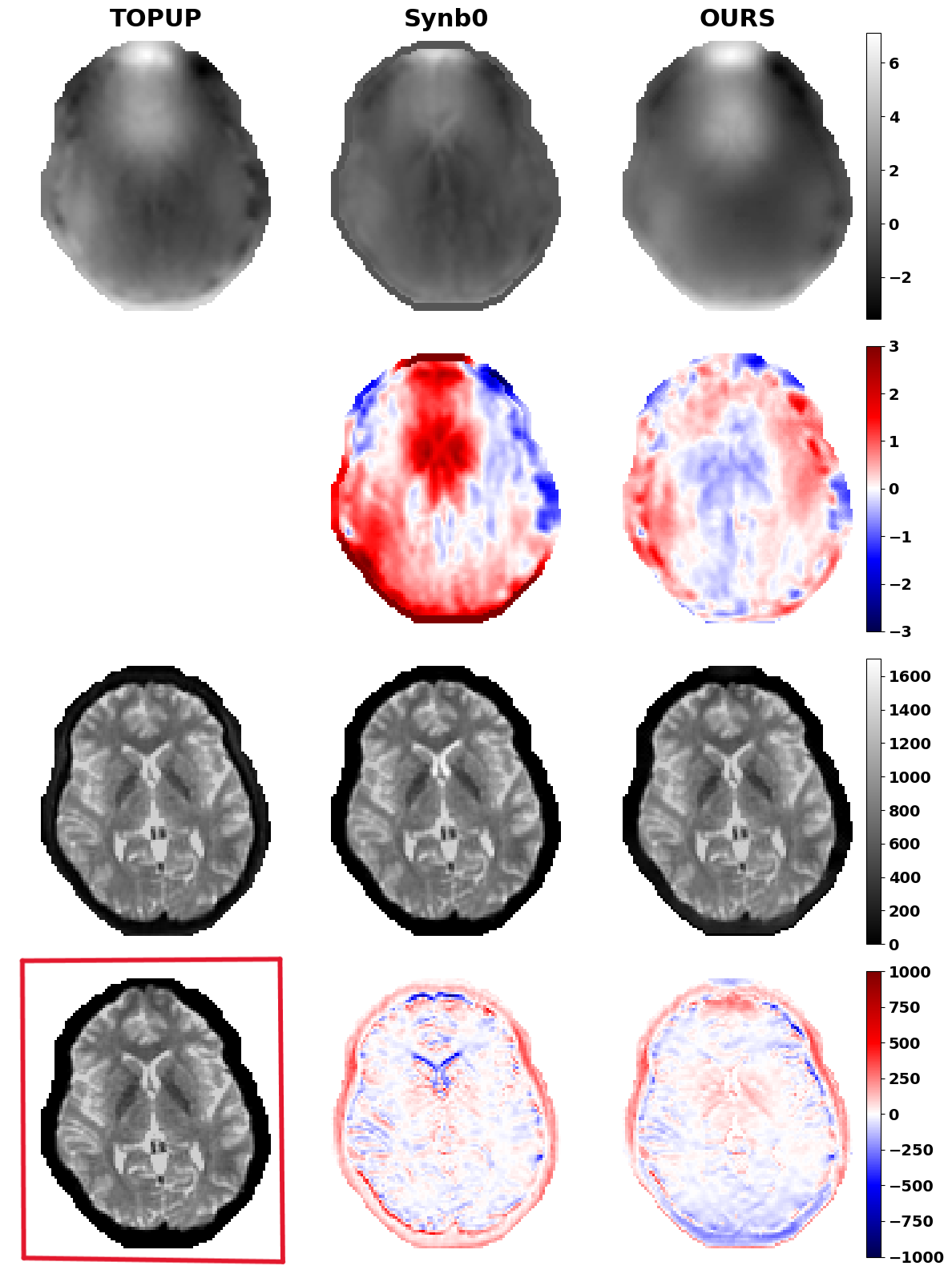}}
  \end{minipage}
  \caption{Comparison of \topup{}\index{topup}, Synb0, and our method for two NIMH-HV test subjects: (left) ON94856 slice 21 and (right) ON95003 slice 36. Rows show: predicted VDMs\index{voxel displacement map (VDM)}, VDM\index{voxel displacement map (VDM)} differences vs. \topup{}\index{topup}, corrected b0 images, and distorted dMRI\index{dMRI} (with red box) alongside b0 differences.}
  \label{fig:side_by_side_dif}
\end{figure}

Figure \ref{fig:side_by_side_dif} showcases the results of the proposed method, including the predicted VDMs\index{voxel displacement map (VDM)} and corrected b0 images, compared to those generated by \topup{}\index{topup} and Synb0. While \topup{}\index{topup} typically achieves the best anatomical alignment, thanks to its use of both blip-up\index{blip-up} and blip-down\index{blip-down} acquisitions, it cannot be applied when only a single phase-encoded\index{Phase Encoding} image is available. Synb0 often under or overestimates broad distortion patterns. By contrast, our network learns a more accurate displacement map from a single input image. These results suggest that our approach can deliver near–\topup{}\index{topup} quality with a single input image, making it a versatile alternative in data-limited scenarios.

\begin{figure}[!t]
\centering
\includegraphics[width=0.65\textwidth]{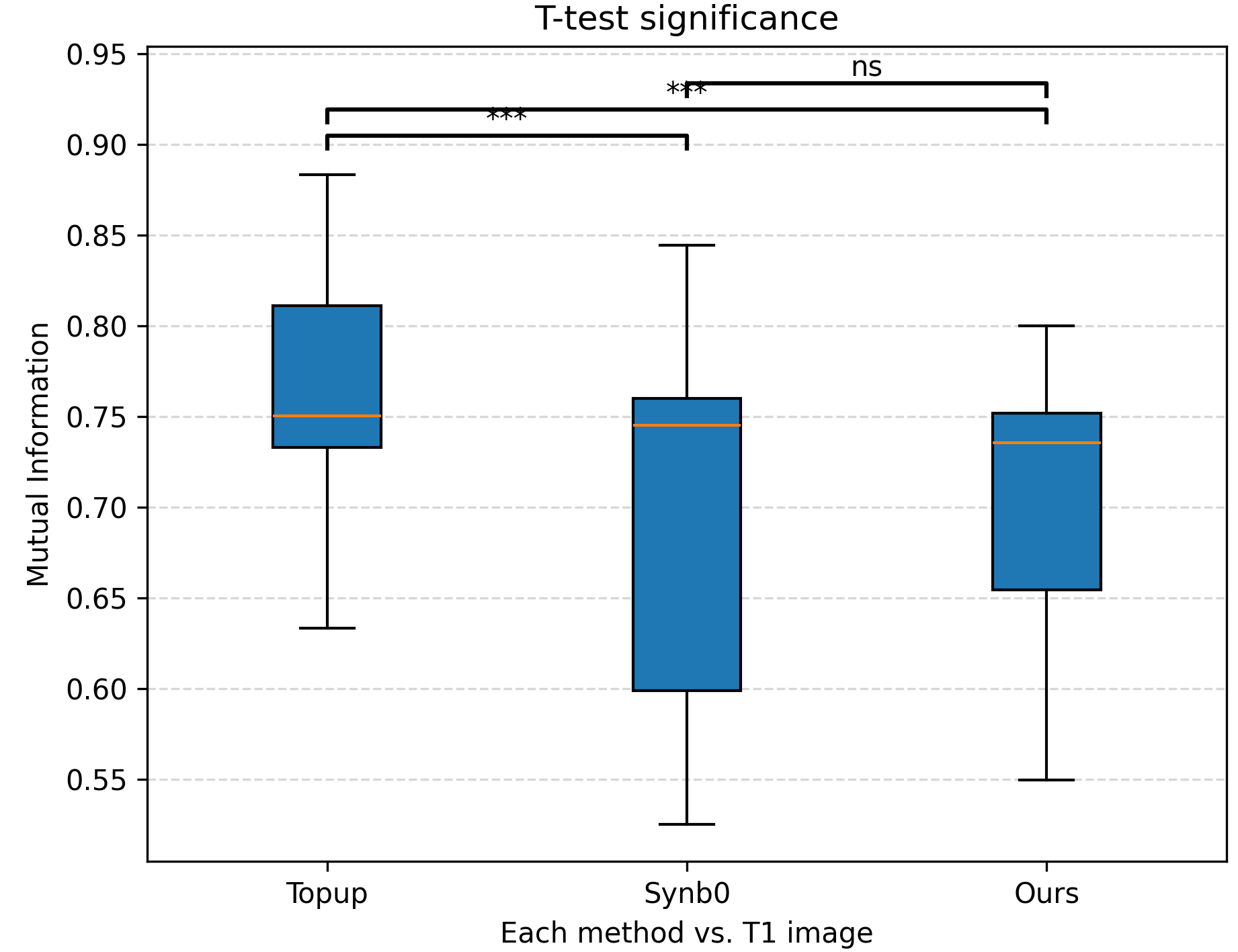}
        \caption{Each method's MI against the T1-weighted image for paired t-test.}  \label{fig4}
\end{figure}

We also compared each method's MI against the T1w image using paired t-test (you can see the results in Figure \ref{fig4}). Under the paired t-test, \topup{}\index{topup} achieves the highest median MI and outperforms Synb0 as well as our method, whereas the gap between Synb0 and ours does not reach significance. These statistics show that our approach achieves anatomical alignment on par with Synb0 and nearly as good as \topup{}\index{topup}, despite using only one encoding direction.

\subsection{Ablation study}
In this section, we evaluate two important design choices we made in building our network: (\emph{i}) the use of a T1w image as input to guide anatomical accuracy and (\emph{ii}) the addition of the gradient of the VDM\index{voxel displacement map (VDM)} as a loss term to improve sharpness in VDM\index{voxel displacement map (VDM)} reconstruction.

\mypar{Using T1w.} In this experiment, we evaluate how much the T1w image contributes when used both as an additional input and as a term in the loss function. To isolate its effect, we retrain our model using only the b0 image as an input and set the weight of the T1w loss term to zero. Table \ref{tab:tab3} summarizes the results. As can be seen, including the T1w image leads to an improvement across every metric. The VDM\index{voxel displacement map (VDM)} RMSE decreases by nearly 26\%, the corrected b0 error drops by roughly 11\%, and MI improves by about 7\%, indicating that the model learns more precise displacement fields when guided by T1w's distortion-free anatomy.  This confirms that the T1w image provides valuable anatomical context that improves correction quality.

\begin{figure}[!t]
  \centering
  \begin{minipage}[b]{0.48\textwidth}
    \centering
    \fbox{\includegraphics[width=\textwidth]{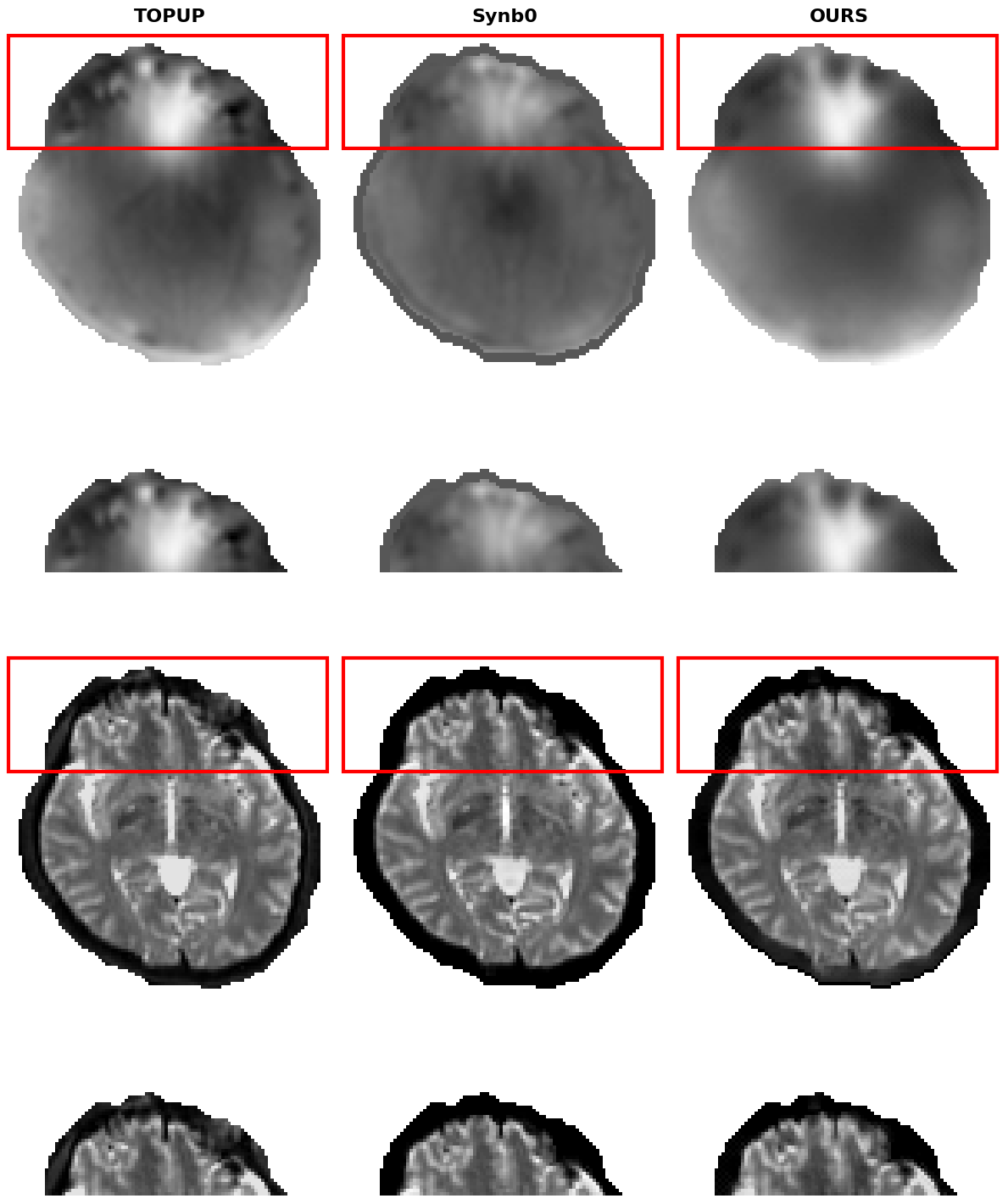}}
  \end{minipage}
  \hfill
  \begin{minipage}[b]{0.48\textwidth}
    \centering
    \fbox{\includegraphics[width=\textwidth]{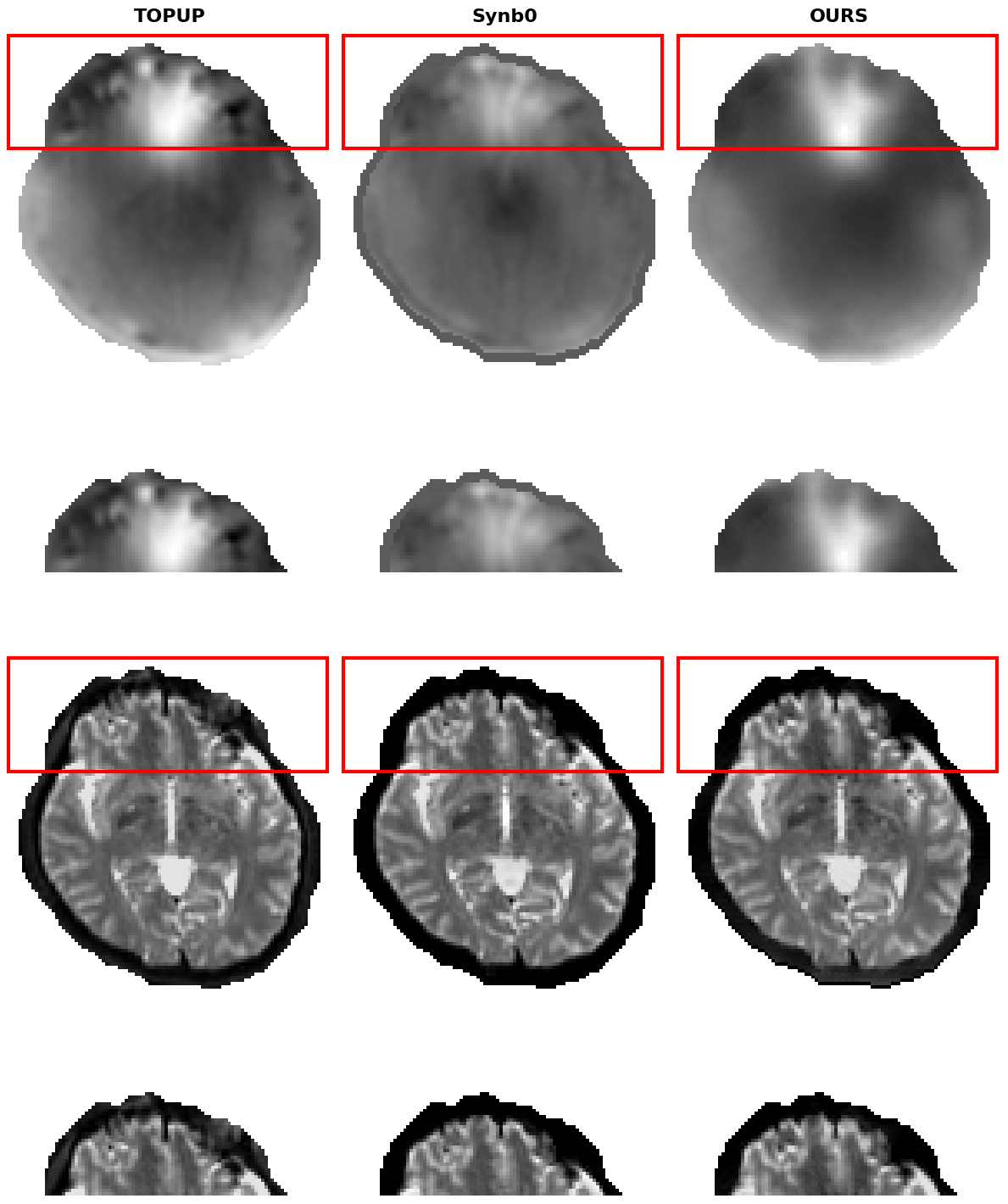}}
  \end{minipage}
  \caption{Visual comparison of \topup{}\index{topup}, Synb0, and our method for NIMH-HV subject ON93426 (slice 36) left image with and right image without the VDM\index{voxel displacement map (VDM)} gradient loss. The results indicate minimal visual differences between models trained with or without the gradient loss term, highlighting the network's inherent ability to produce smooth displacement fields.}
  \label{fig:side_by_side_ablation}
  
\end{figure}

\mypar{Using the gradient loss term.} Next, we remove the gradient‐based term on the predicted VDMs\index{voxel displacement map (VDM)} to understand its impact on the final correction. As reported in the bottom half of Table~\ref{tab:tab3}, removing the gradient term leads to a modest increase in VDM\index{voxel displacement map (VDM)} and b0 RMSE, a slight raise in MI with the T1w image. Both sets of VDMs\index{voxel displacement map (VDM)}, whether trained with or without the gradient penalty, remain smoother than the \topup{}\index{topup} reference as shown in Figure \ref{fig:side_by_side_ablation}, suggesting our network's architecture already favors smooth displacement estimates. Therefore, the extra gradient‐loss term delivers only a small numeric gain in RMSE but does not yield any clear boost in anatomical alignment.

\begin{table}[!t]
\centering
\caption{Effect of using T1w image and including the VDM\index{voxel displacement map (VDM)} gradient loss on performance. All metrics are averaged across the held-out test subjects and reported as mean (standard deviation).}
\label{tab:tab3}
\setlength{\tabcolsep}{0.5em}
\resizebox{.92\linewidth}{!}{
\begin{tabular}{l|c|c|c}
\toprule
\textbf{Configuration} & \textbf{VDM RMSE}\,$\downarrow$ & \textbf{b0 RMSE}\,$\downarrow$ & \textbf{MI}\,$\uparrow$ \\
\midrule
With T1w    &  1.10 (0.31)     &  $1.76 \times 10^2$ (0.339$\times 10^2$)   & 0.7062 (0.0743) \\
Without T1w & 1.48 (0.49)   &  $1.97 \times 10^2$ (0.422$\times 10^2$) & 0.6605 (0.0698) \\
\midrule
With gradient loss          &  1.10 (0.31)     &  $1.76 \times 10^2$ (0.339$\times 10^2$)   & 0.7062 (0.0743)                   \\
Without gradient loss       & 1.22 (0.35)             & $1.80\times10^{2}$ (0.344$\times10^{2}$)  & 0.7080 (0.0742)                  \\
\bottomrule
\end{tabular}
}
\end{table}

\subsection{Conclusion}

In this work, we introduce a deep learning–based\index{deep learning} framework for susceptibility distortion\index{susceptibility distortion} correction in diffusion MRI\index{Diffusion MRI} that requires only a single blip-up\index{blip-up} or blip-down\index{blip-down} acquisition. By training a 2.5D UNet to jointly predict the VDM\index{voxel displacement map (VDM)} and the intensity-corrected b0 image, we demonstrate that it is possible to recover nearly the same geometric and contrast information that \topup{}\index{topup} provides in a faster way, despite never seeing a reverse-phase image.  Across both the NIMH-HV dataset and unseen CS-DSI samples, our method is more accurate and significantly faster than Synb0.

Nevertheless, a performance gap remains between our single-input correction and the dual-phase silver standard \topup{}\index{topup}. In future work, we aim to further close this gap by refining our model and extending it to address additional artifacts such as eddy currents and subject motion. We also plan to investigate the model’s applicability to other PE directions, and quantify downstream effects for example, FOD coherence, tractography accuracy, and scan–rescan reproducibility. Overall, our findings highlight the potential of deep learning\index{deep learning} as a practical, fast, and versatile alternative to traditional correction methods in diffusion MRI\index{Diffusion MRI}; especially in scenarios where only a single phase-encoding\index{Phase Encoding} direction is available.

\begin{credits}
\subsubsection{\ackname} This research is supported by NIMH award U24MH124629 (SD, SB) and the Canada Research Chairs Program (SB). 
\subsubsection{\discintname} We do not have any competing interests.

\end{credits}
%
%
%
%

\end{document}